\documentclass[twocolumn,
aps,prb,twoside,superscriptaddress,floatfix,tightenlines,nofootinbib]{revtex4-1}
\usepackage{dcolumn}
\usepackage{float}
\usepackage{bm}
\usepackage[T1]{fontenc}
\usepackage{amsmath}
\usepackage{graphicx} 
\usepackage{tabularx}
\usepackage{multirow}
\usepackage{graphics}
\usepackage[version=4]{mhchem}
\usepackage[retainplus]{siunitx}
\usepackage{rotating} 
\usepackage{makecell}
\usepackage{hyperref}
\hypersetup{colorlinks=true, linkcolor=blue, citecolor=blue, urlcolor=blue}
\usepackage{tikz}
\usepackage[normalem]{ulem}
\usepackage[utf8]{inputenc}
\usepackage{booktabs}
\usepackage{mathtools}
\usepackage{siunitx}

\DeclarePairedDelimiterXPP\BigOSI[2]%
  {\mathcal{O}}{(}{)}{}%
  {\SI{#1}{#2}}

\begin{document}

\title{Prediction of Tunable Spin-Orbit Gapped Materials for Dark Matter Detection}

\author{Katherine Inzani}
\email{kinzani@lbl.gov}
\affiliation{Materials Science Division, Lawrence Berkeley National Laboratory, Berkeley, CA 94720, USA}
\affiliation{Molecular Foundry, Lawrence Berkeley National Laboratory, Berkeley, CA 94720, USA}
\author{Alireza Faghaninia}
\affiliation{Energy Technologies Area, Lawrence Berkeley National Laboratory, Berkeley, CA 94720, USA}

\author{Sin\'{e}ad M. Griffin}
\affiliation{Materials Science Division, Lawrence Berkeley National Laboratory, Berkeley, CA 94720, USA}
\affiliation{Molecular Foundry, Lawrence Berkeley National Laboratory, Berkeley, CA 94720, USA}

\date{\today}

\begin{abstract}
New ideas for low-mass dark matter direct detection suggest that narrow band gap materials, such as Dirac semiconductors, are sensitive to the absorption of meV dark matter or the scattering of keV dark matter. Here we propose spin-orbit semiconductors -- materials whose band gap arises due to spin-orbit coupling -- as low-mass dark matter targets owing to their $\BigOSI{10}{meV}$ band gaps. We present three material families that are predicted to be spin-orbit semiconductors using Density Functional Theory (DFT), assess their electronic and topological features, and evaluate their use as low-mass dark matter targets. In particular, we find that that the tin pnictide compounds are especially suitable having a tunable range of meV-scale band gaps with anisotropic Fermi velocities allowing directional detection. Finally, we address the pitfalls in the DFT methods that must be considered in the \textit{ab initio} prediction of narrow-gapped materials, including those close to the topological critical point.

\end{abstract}

\maketitle

\section{Introduction}

New models of dark matter (DM) offer the tantalizing possibility that direct detection is within the realms of short and modestly scaled experiments.\cite{cosmic2017a} Recent models assigning DM mass to the sub-GeV range have incentivized the design of detection experiments that push the bounds of mass sensitivity. The observation of the small energy depositions associated with light masses requires creative materials solutions, with recently proposed targets including scintillators, Dirac materials, superconductors, polar materials and superfluid helium.\cite{Derenzo2017a,Hochberg2018,Hochberg2016a,Hochberg2016b,Knapen2018a,Knapen2017a,Trickle2019,Griffin2019}

Charge-based detectors rely on scattering or absorption events to excite charge carriers across an energy gap that is tailored to the expected energy deposition.\cite{Hochberg2017a,Essig2016} Semiconductors with meV-scale band gaps are therefore suitable for the absorption of DM with meV mass and scattering of DM with keV mass due to the meV magnitude kinetic energy. Although the energy gap imposes a threshold on the detectable DM mass, a finite gap is necessary for decoupling a DM signal from thermal noise.\cite{Hochberg2016b} Therefore, semiconductors with ultra-narrow band gaps are sought to maximize the reach of direct detection experiments. Narrow band gap semiconductors are also desired for infrared radiation detection, especially for sensitivity to long wavelengths.\cite{Baker2017} Furthermore, small band gaps are often linked to high-performance of conventional thermoelectric materials.\cite{Shi2016}

A special class of small band gap compounds, (gapped) Dirac materials, have been identified as promising DM detection targets providing high sensitivity for absorption events.\cite{Hochberg2018} For maximal DM scattering rate, the target material should have a Fermi velocity kinematically matched to the DM velocity, which is serendipitously of the order of $10^5$ $\mathrm{ms^{-1}}$, similar to the ranges reported in Dirac materials.\cite{Hochberg2018}

Several candidates for low-mass DM detection based on low band gap, Dirac-like dispersions have been explored,\cite{Geilhufe_et_al:2018, Geilhufe_et_al:2020, Sanchez_et_al:2019} with topological ZrTe$_5$ emerging as a leading candidate.\cite{Hochberg2018} ZrTe$_5$ possesses Dirac nodes that are gapped out to $\sim$20 meV when spin-orbit coupling (SOC) is included in calculations.\cite{Nair_et_al:2018} However, ZrTe$_5$ is difficult to obtain in large single-crystal form as it is a layered van der Waals material, and in addition, its electronic properties have been shown to strongly depend on structure, synthesis conditions and temperature.\cite{Xu_et_al:2018,Monserrat_et_al:2019} Therefore, alternative low-band gap materials which can be synthesized reliably in single-crystal form are needed for next-generation low mass dark matter experiments.

However, any computational searches for such low band gap materials will be mired with the well-documented band gap problem of standard Density Functional Theory (DFT) methods. Semilocal DFT exchange-correlation functionals include a spurious self-interaction in the occupied states, resulting in the over-delocalization of charge densities; in addition, semilocal functionals do not feature a discontinuity in the potential with change in particle number, resulting in significant underestimation of band gaps.\cite{Perdew1983,Perdew2009,Mori-Sanchez2008a} In many materials the electron delocalization is better treated with a screened hybrid functional, which can improve the description due to a reduction in the self-interaction error, and open the band gap.\cite{Krukau2006a} Hybrid functionals have a greater computational cost than semilocal DFT, but are considerably cheaper than the more chemically accurate GW methods, whilst providing comparable results.\cite{Garza2016,Crowley2016}

In this work we propose an alternative method to circumvent this common failure of DFT in band gap prediction. Here we propose to search instead for `spin-orbit semiconductors' -- these would-be metals are metallic without SOC and are gapped out upon the inclusion of SOC. A familiar example occurs in graphene which is a Dirac semimetal when SOC is not included, opening up to a gap of tens of $\mu$eV with SOC. In fact, the concept of a spin-orbit gap, that is a band gap that is opened only when SOC interactions are included, is closely related to the topological quantum phase transition, whereby a topological material, such as a Dirac semimetal, can become a trivial insulator by manipulation of the symmetry of the crystal potential. Such spin-orbit gaps can be predicted by DFT calculations by simply comparing band structures with SOC included and not included. Band gaps are of the order of the strength of spin-orbit coupling, hence in the $\sim$meV range suitable for low-mass dark matter detector candidates.

In this work, we examine the use of a spin-orbit gap, as predicted by DFT, for selecting candidate materials for dark matter detection targets. We choose three materials  predicted to have a spin-orbit gap and evaluate their electronic structure, using generalized gradient approximation (GGA) and hybrid functional levels of theory, and their closeness to the topological point. Following the computational methodology, the results and discussion are divided into three sections: A. Candidate Materials, B. Suitability as Dark Matter Targets and C. Theoretical Predictions of Spin-Orbit Gapped Materials.

\section{Computational Methodology}

First-principles calculations based on Density Functional Theory (DFT) were performed using the Vienna \textit{Ab initio} Simulation Package (\textsc{vasp})\cite{Kresse1993,Kresse1994,Kresse1996,Kresse1996a} with projector augmented wave (PAW) pseudopotentials.\cite{Blo,Kresse1999} For each element, the states included as valence were \textit{s} and \textit{p} for alkali metals, alkaline earth metals, metalloids and non-metals, and \textit{s}, \textit{p} and \textit{d} for transition and post-transition metals. Energy cutoff and k-point convergence testing was carried out on each material and parameters were chosen for a convergence of at least 1 meV per formula unit. These parameters are given in SI Table 1. The convergence criteria for the electronic self-consistent loop was set to $10^{-7}$ eV. Structural optimizations were done using the Perdew-Becke-Ernzerhof (PBE)\cite{Perdew1996} exchange-correlation functional until the residual forces on the ions were less than \SI{0.001}{\electronvolt\per\angstrom}.

Electronic density of states and band structures were calculated with both the PBE functional and the HSE06 hybrid functional\cite{Heyd2003a,Heyd2006} on top of the PBE optimized structures. Band structures were calculated both with and without spin-orbit coupling interactions, which were included self-consistently.\cite{Steiner2016} The software \textsc{sumo} was used to plot the electronic structures.\cite{Ganose2018}

Topological characterization was carried out using \textsc{symtopo}.\cite{Symtopo} The topological invariant Z\textsubscript{2} and the surface states were calculated using WannierTools\cite{WU2017} with Wannier centres calculated using Wannier90.\cite{Pizzi2020}

\section{Results and discussion}

\begin{figure}[!tbh]
 \centering
 \includegraphics[width=\columnwidth]{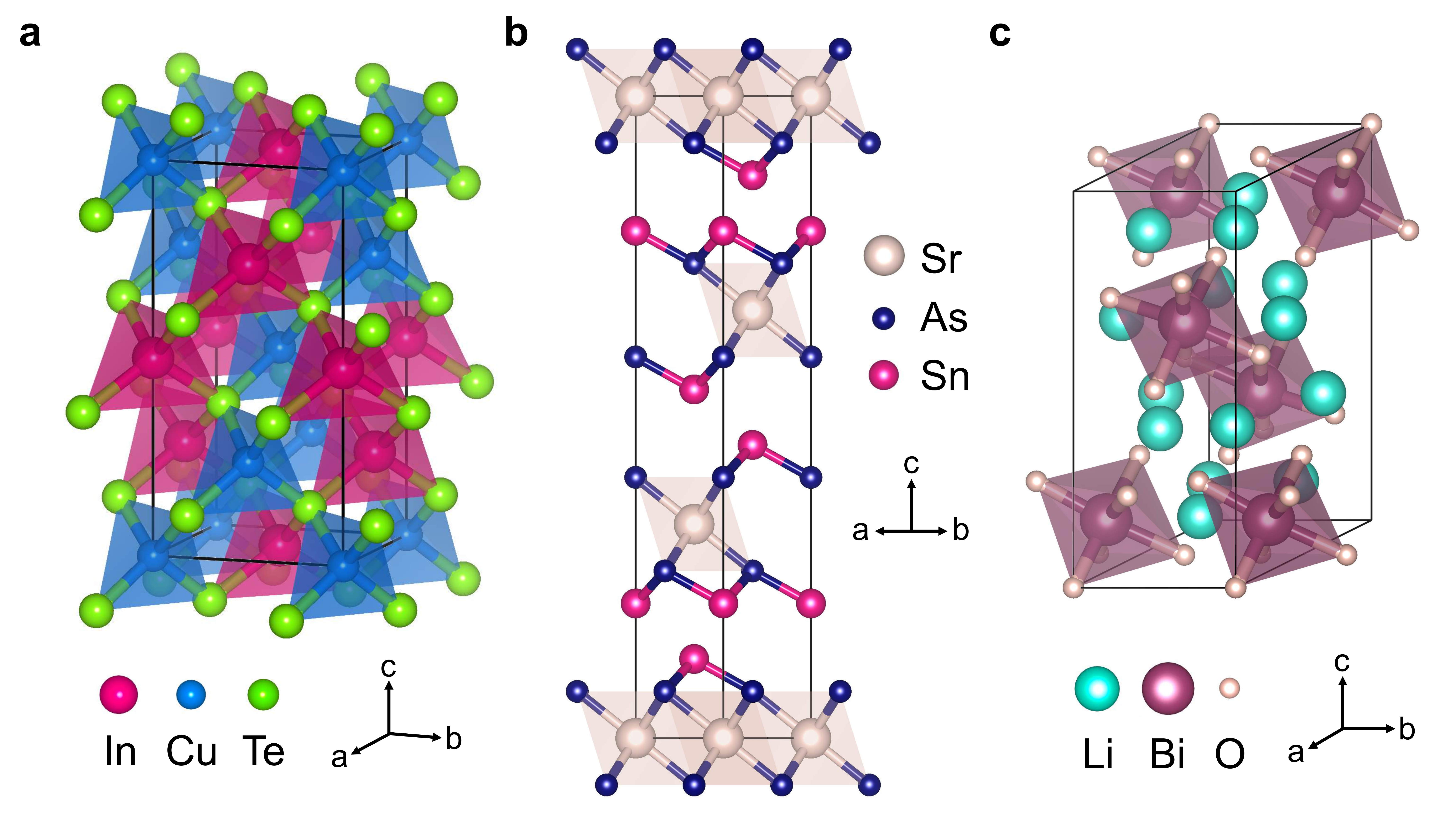}
 \caption{Crystal structures of three materials gapped by spin-orbit coupling, \textbf{a} \ce{CuInTe2}, \textbf{b} \ce{SrSn2As2} and \textbf{c} \ce{Li6Bi2O7}.}
 \label{structures}
\end{figure}

\subsection{Candidate Materials}

We used an existing data set of materials\cite{private} which had electronic transport properties calculated by DFT, using the PBE functional, to select three candidate materials which were metallic (zero gap) without including SOC, and opened up a finite gap with SOC included: \ce{CuInTe2}, \ce{SrSn2As2} and \ce{Li6Bi2O7}. We ensured that these compounds contained at least one element with a sizable spin-orbit coupling magnitude and had a calculated energy above hull less than 0.02 eV. We also investigated materials which were isostructural to these three candidates, with cation and anion substitutions chosen to vary the spin-orbit interaction strength. In a spin-orbit gapped material, substitutions of heavier elements would be expected to widen the band gap due to the increase in SOC-strength with atomic mass.\cite{Herman1963} We report the trends in electronic structure with these substitutions and their effectiveness in tuning the band gap and hence sensitivity to various DM masses.

\subsubsection{Copper indium chalcogenides}
\noindent
\textbf{Structural details}: Copper indium ditelluride, \ce{CuInTe2}, has the chalcopyrite crystal structure (space group I$\overline{4}$2d, number 122), shown in Figure \ref{structures}a, which can be described as the zinc blende structure doubled in the \textit{c}-direction due to the alternating Cu\textsuperscript{+} and In\textsuperscript{3+} sites. Each Cu and In is tetrahedrally coordinated with Te, forming a checkerboard corner-sharing network. Substituting Se or S onto the Te sites yields \ce{CuInSe2} and \ce{CuInS2} which are also known to exist in the chalcopyrite structure.\cite{Rincon1992,Hwang1978} The calculated lattice parameters of the three copper indium chalcogenides are well matched to reported experimental values (given in SI Table 2). The chalcopyrite family have previously been considered for light-harvesting devices including solar cells, solar fuel cells and photodetectors,\cite{Kazmerski1977,Neumann1986,Rockett1991,Yoshino2001,Bi2012,Frick2018c} and \ce{CuInTe2} has also been suggested as a promising thermoelectric.\cite{Liu2012b} \medskip

\begin{figure}[!tb]
 \centering
 \includegraphics[width=\columnwidth]{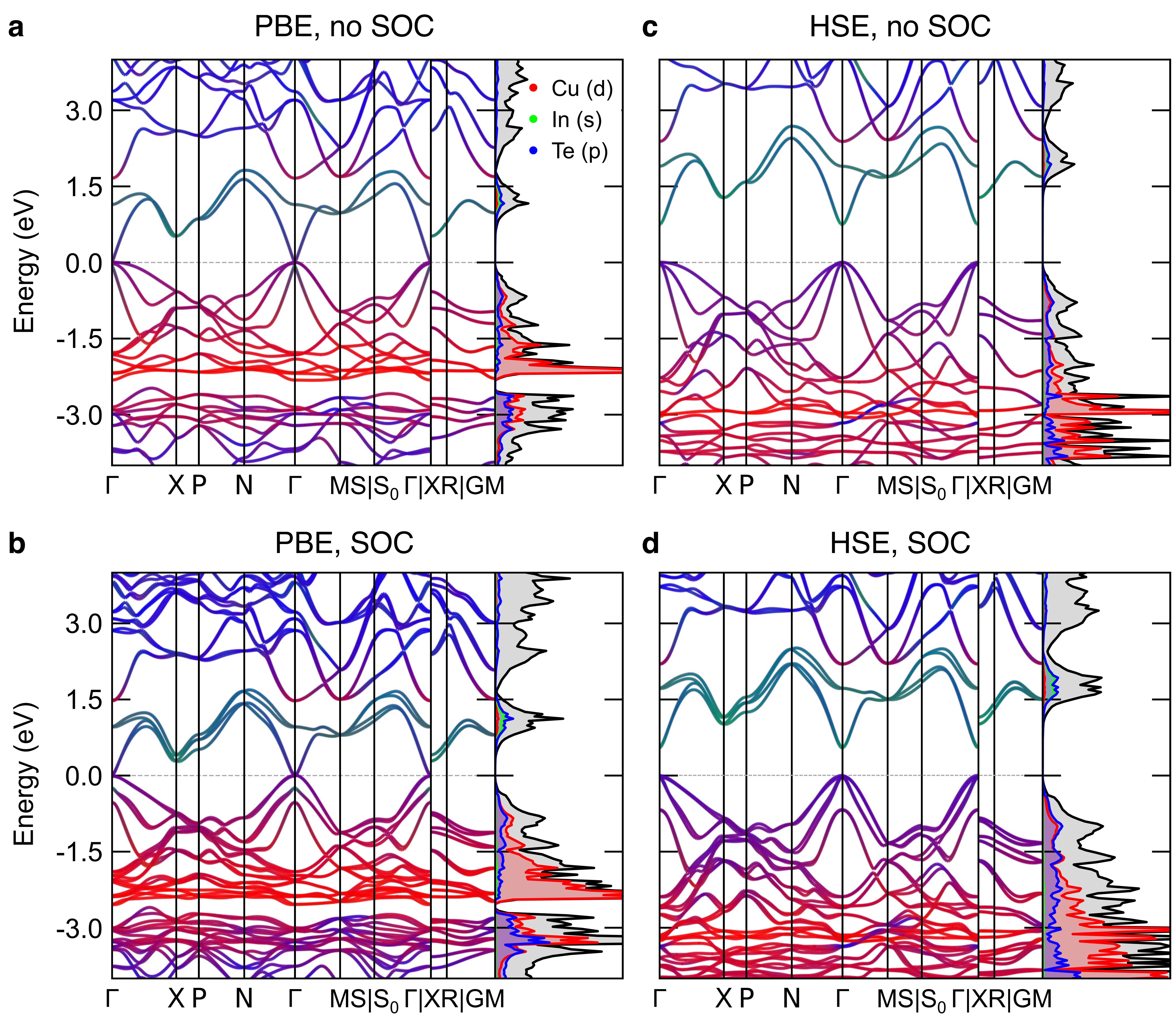}
 \caption{\textbf{a-b:} Electronic band structures and orbital-resolved density of states of \ce{CuInTe2} calculated with PBE: \textbf{a} without spin-orbit coupling included and \textbf{b} with spin-orbit coupling included. \textbf{c-d:} Calculation with HSE06 results in a large band gap opening in both \textbf{c} without spin-orbit coupling included and \textbf{d} with spin-orbit coupling included. The orbitals Cu 3\textit{d} (red), In 5\textit{s} (green) and Te 5\textit{p} (blue) are projected onto the bands.}
 \label{cuinte2_bs}
\end{figure}

\begin{figure*}[!tb]
 \centering
 \includegraphics[width=12cm]{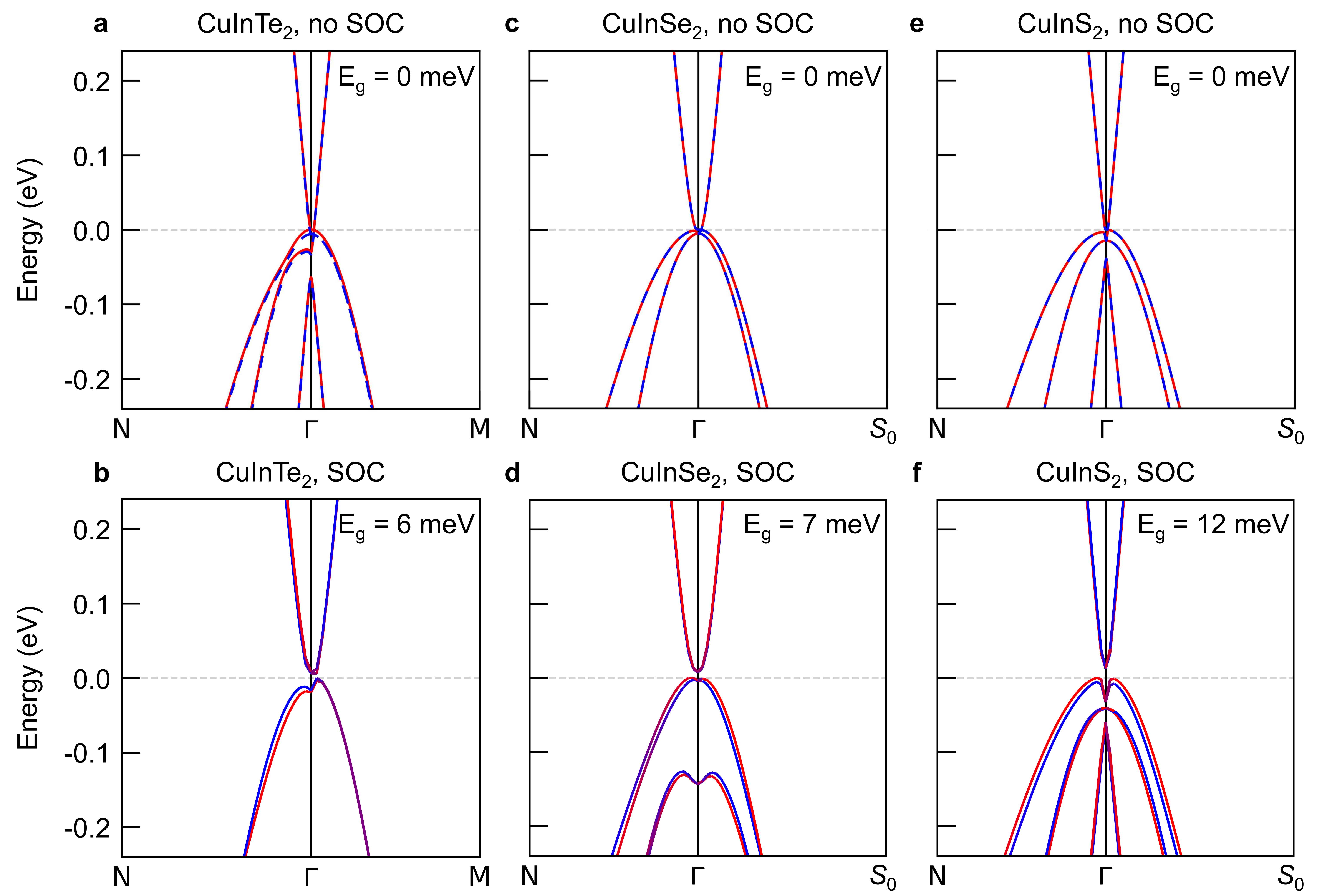}
 \caption{Magnified, spin-projected band structures of \ce{CuInTe2}, \ce{CuInSe2} and \ce{CuInS2} calculated with PBE. The top row is without spin-orbit coupling included, where red and blue correspond to spin-up and spin-down channels. The bottom row is with spin-orbit coupling included, where red and blue show the spin-projection onto the \textit{x}-direction. With spin-orbit coupling included, the valence band maximum of \ce{CuInTe2} sits along the $\Gamma$ to M direction, and for \ce{CuInSe2} and \ce{CuInS2} along the N to $\Gamma$ direction.}
 \label{cuin_zoom_bs}
\end{figure*}

\noindent
\textbf{Electronic structure}: The electronic band structures and density of states (DOS) of \ce{CuInTe2} as calculated with the PBE functional with SOC both included and not included are shown in Figure \ref{cuinte2_bs}a-b. The orbital-resolved DOS reveals the valence band to be mainly Cu 3\textit{d} and Te 5\textit{p} character, while the conduction band contains a fraction of In 5\textit{s} states, as shown in the orbital projections in Figure \ref{cuinte2_bs}. With SOC not included, at $\Gamma$ there are two doubly-degenerate bands at the Fermi level and two doubly-degenerate bands very close to the Fermi level, as shown in Figure \ref{cuin_zoom_bs}a. The effect of SOC is to lift these degeneracies and introduce further spin splitting throughout the band structure. In addition, the band gap (E\textsubscript{g}) with PBE, which is zero when SOC is not included, is opened to 6 meV with SOC. This is more clearly seen in Figure \ref{cuin_zoom_bs}a-b, which shows the band structure magnified around the Fermi level at $\Gamma$ and projected onto the different spin channels. For the band structures without SOC included, the spin-up and spin-down channels are degenerate. When SOC is included, the spin projection onto the \textit{x}-direction reveals the degeneracy to be lifted by spin component. This Dresselhaus spin splitting results from the lack of inversion symmetry in the structure of \ce{CuInTe2} and presence of spin-orbit coupling. In fact, the Dresselhaus effect was originally proposed for zinc blende structures from which the chalcopyrite structure of \ce{CuInTe2} is derived.\cite{Dresselhaus1955}

For the \ce{CuInSe2} and \ce{CuInS2} structures, E\textsubscript{g} is also zero when SOC is not included, and E\textsubscript{g} is opened to 7 meV and 12 meV respectively when SOC is included, also shown in Figure \ref{cuin_zoom_bs}. Different band structure paths are shown due to the position of the valence band maximum (VBM) and conduction band minimum (CBM): for \ce{CuInTe2} with SOC, the VBM and CBM are slightly away from $\Gamma$ in the $\Gamma$ $\rightarrow$ M direction, whereas \ce{CuInSe2} and \ce{CuInS2} have the VBM along $\Gamma$ $\rightarrow$ $S_{0}$ and the CBM  at $\Gamma$ with no SOC, and the VBM along $\Gamma$ $\rightarrow$ N and the CBM  at $\Gamma$ with SOC included. Contrary to what is expected for SOC-mediated band inversion where heavier ions would result in greater E\textsubscript{g}, here we find that E\textsubscript{g}  decreases with increasing anion mass (S$\rightarrow$Se$\rightarrow$Te). In this case, the E\textsubscript{g} differences are a result of  higher energy \textit{p}-orbitals hybridizing more strongly with metal \textit{d}-orbitals and forming a more disperse valence band. Furthermore there is no evidence of band inversion seen in the orbital projections on the bands given in SI Figure 1. However, the enhanced spin-splitting effect is clearly seen in the spin-projected bands (Figure \ref{cuin_zoom_bs}) and in comparison of the full band structures of the materials series (SI Figure 1).

The generalized gradient approximation on which the PBE functional is based is well-known to underestimate band gaps. Therefore, the band structures were also calculated with the hybrid functional HSE06, which has been shown to perform better for E\textsubscript{g} in semiconductors.\cite{Krukau2006a} E\textsubscript{g} of \ce{CuInTe2} is opened to 0.74 eV without SOC, and the effect of SOC is much larger - reducing E\textsubscript{g} to 0.54 eV as shown in Figure \ref{cuinte2_bs}c-d. The band gaps of \ce{CuInSe2} and \ce{CuInS2} without (and with) SOC included are 0.62 eV (0.56 eV) and 1.08 eV (1.08 eV) respectively. The differences in band gaps show the diminishing effect of SOC through the series as lighter elements have smaller relativistic effects.\cite{Herman1963} This can also be seen in the reduction in spin-splitting across the band structures calculated with HSE06, which are given in SI Figure 2.

\subsubsection{Tin pnictides}
\noindent
\textbf{Structural details}: Next we turn to \ce{SrSn2As2}, shown in Figure \ref{structures}b, which belongs to the tin pnictide family (space group R$\overline{3}$m, number 166). The crystal structure consists of layers of edge-sharing \ce{SrAs6} octahedra alternating with layers of Sn. The tin pnictides recently gained interest when \ce{NaSn2As2} and Na\textsubscript{1-x}Sn\textsubscript{2}P\textsubscript{2} were found to be superconductors with \textit{T\textsubscript{C}} $\sim$1.3 K and 2.0 K respectively.\cite{Goto2017,Goto2018} Unlike the exotic superconductivity present in the stoichiometrically similar Fe-pnictide compounds (e.g. BaFe$_2$As$_2$) with the Fmmm space group, \ce{NaSn2As2} has been classified as a phonon-mediated conventional superconductor.\cite{Ishihara2018}

The tin pnictides are also isostructural to the topological insulator \ce{Bi2Te2Se}, consisting of layers of \ce{SeBi6} octahedra alternating with layers of Te.\cite{Xiong2012,Bland1961} \ce{SrSn2As2} itself has been theoretically predicted to be an enforced 3D-Dirac semimetal lying naturally close to the topological critical point.\cite{Gibson2015} Experimental evidence of a topological insulating state in \ce{SrSn2As2} from angle-resolved photoemission spectroscopy has been reported by Rong \textit{et al.}, although interpretation of these results was not clear-cut and the band gap of the structure was not measured.\cite{Rong2017} They also note from their DFT calculations that the topological state was sensitive to the choice of exchange-correlation functional.

To investigate the role of SOC and ion sizes, we fully substituted isovalent ions on both the Sr and As sites. For the Sr site, we considered other alkali earth metals with \ce{MgSn2As2}, \ce{CaSn2As2} and \ce{BaSn2As2}. For the As site we substituted in P, resulting in \ce{SrSn2P2}. The optimized calculated lattice parameters are compared to experimental values in SI Table 3. There are limited measurements of the lattice parameters reported for these compounds, and in some cases we have estimated the lattice parameter by extrapolating from the available experimental lattice parameters of mixed cation compounds according to Vegard's law. Taking this into consideration, the calculated lattice parameters compare favourably with reported structures.\cite{Asbrand1995} Both lattice parameters \textit{a} and \textit{c} increase with increasing cation mass (Mg$\rightarrow$Ca$\rightarrow$Sr$\rightarrow$Ba) and anion mass (P$\rightarrow$As). \medskip

\begin{figure*}[!tb]
 \centering
 \includegraphics[width=18cm]{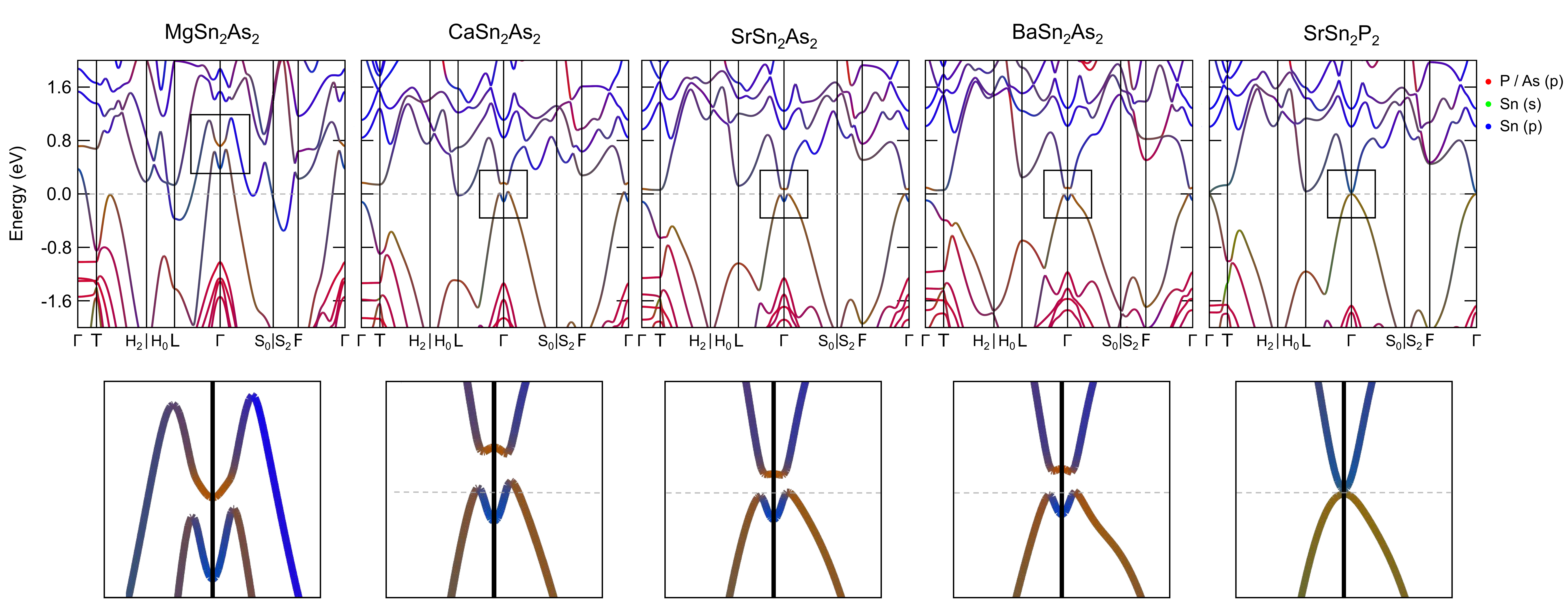}
 \caption{Orbital projected band structures of \ce{MgSn2As2}, \ce{CaSn2As2}, \ce{SrSn2As2}, \ce{BaSn2As2} and \ce{SrSn2P2} calculated by PBE with spin-orbit coupling included showing the Sn \textit{s} (green), Sn \textit{p} (blue) and P/As \textit{p} (red) orbitals. The bottom panel shows magnifications of the band inversion near the Fermi level in each case. For all plots, the Fermi level is set to 0 eV. }
 \label{tin_gga_ncl_bs}
\end{figure*}
 
\begin{table}[!]
\centering
\caption{Band gaps (E\textsubscript{g}) of tin pnictides calculated by PBE and HSE06, with and without spin-orbit coupling. Bold font indicates a direct band gap.}
\label{tin_pnictide_Egs}
\begin{tabular}{@{}lcccc@{}}
\toprule
& \multicolumn{2}{c}{E\textsubscript{g} PBE (eV)} & \multicolumn{2}{c}{E\textsubscript{g} HSE06 (eV)} \\
Material        & Without SOC & With SOC & Without SOC & With SOC \\ \midrule
\ce{MgSn2As2} & 0 & 0 & -- & -- \\ 
\ce{CaSn2As2} & 0 & 0 & 0 & 0 \\ 
\ce{SrSn2As2} & 0 & 0.057 & \textbf{0.029} & \textbf{0.070} \\
\ce{BaSn2As2} & 0 & 0.054 & \textbf{0.049} & \textbf{0.112} \\
\ce{SrSn2P2} & 0.028 & \textbf{0.016} & 0.181 & \textbf{0.198} \\ \bottomrule
\end{tabular}
\end{table}

\noindent
\textbf{Electronic structure}: The calculated PBE+SOC orbital projected band structures are shown in Figure \ref{tin_gga_ncl_bs}. All four of the As-compounds exhibit band inversion at the $\Gamma$ high-symmetry point with avoided band-crossings appearing at several places across all five band structures. This indicates a strong degree of spin-orbit interaction, as many of these avoided crossings are not observed, or occur to a lesser degree, in the band structures calculated without SOC (see SI Figure 3). Calculated band gaps are shown in Table \ref{tin_pnictide_Egs}, with \ce{MgSn2As2} and \ce{CaSn2As2} being metallic, \ce{SrSn2As2} and \ce{BaSn2As2} having a small and indirect band gap, and \ce{SrSn2P2} having a small and direct band gap with PBE. The trend in E\textsubscript{g} with ion mass P$<$As is reversed when SOC is or is not included suggesting that E\textsubscript{g} is strongly dependent on SOC in this structure.

The compounds \ce{CaSn2As2}, \ce{SrSn2As2}, \ce{BaSn2As2} and \ce{SrSn2P2} were further investigated by calculations with HSE06. The band gaps are given in Table \ref{tin_pnictide_Egs} and the HSE06+SOC band structures in SI Figure 4. Aside from \ce{CaSn2As2}, which remains a semimetal, we find that with HSE06 the band gap opens up, and when including SOC the band gap opens further. This latter effect is greatest in \ce{BaSn2As2}, which contains the heaviest elements, then \ce{SrSn2As2}, and finally \ce{SrSn2P2}, the lightest compound considered. Despite this, the band gaps calculated by HSE06+SOC remain at the meV order of magnitude. \medskip

\begin{figure}[!tb]
 \centering
 \includegraphics[width=\columnwidth]{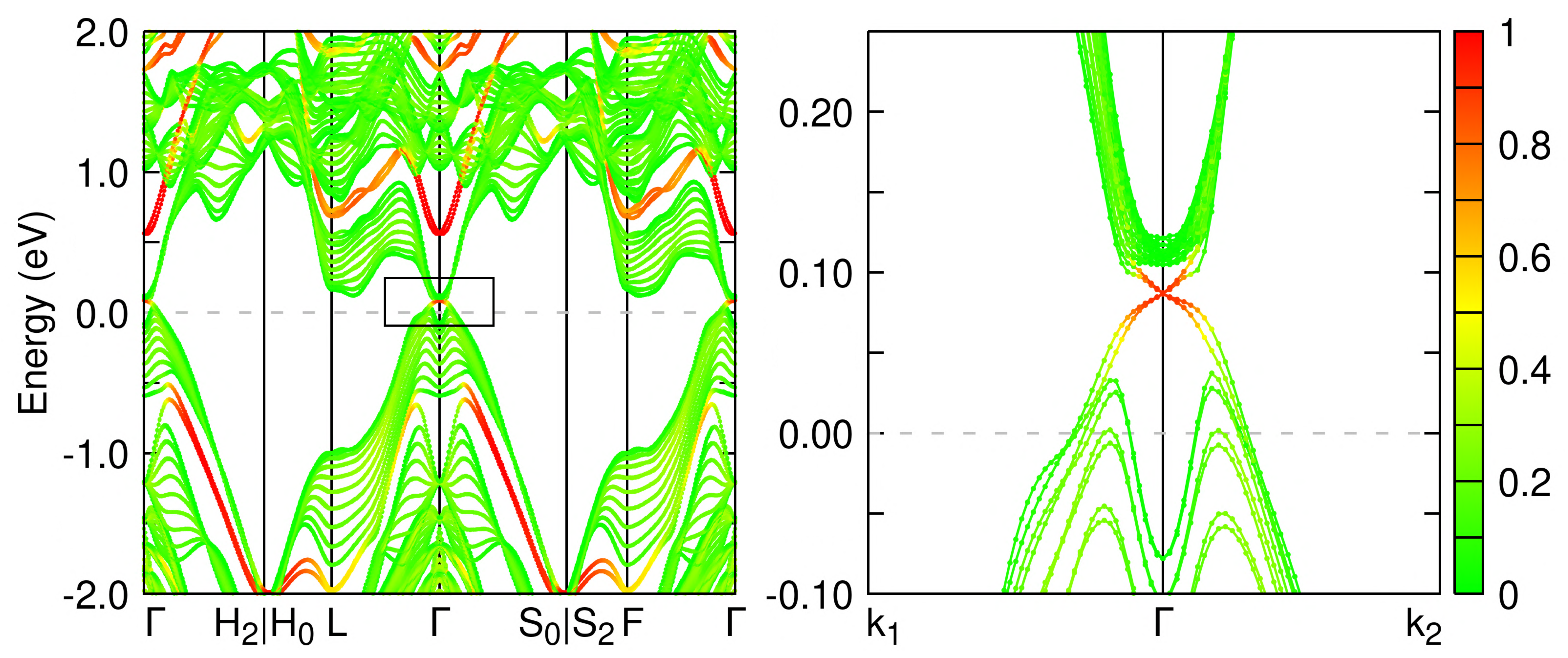}
 \caption{Topological surface states (red) in the band structure of \ce{SrSn2As2} from a 10 layer slab cleaved between Sn layers. The color
scale indicates the weight of projection onto the outermost layers of the slab with green indicating bulk contributions. Right panel is a magnification of the box in the left panel.}
 \label{surface-states}
\end{figure}

Finally, we address the qualitative difference between the band structures at the Fermi level between \ce{MgSn2As2} and the other tin pnictide compounds considered. The orbital-resolved DOS of \ce{SrSn2As2} and \ce{MgSn2As2}, shown in SI Figure 5, reveals states near the top of the valence band and near the bottom of the conduction band to have a mixture of mainly As 4\textit{p}, Sn 5\textit{s} and Sn 5\textit{p} character. This indicates hybridization between Sn and As orbitals within the SnAs bilayers, as is also reported in other tin arsenide layered compounds.\cite{Arguilla2016,Arguilla2017} However, in the Mg case, there are also Mg \textit{s} states at the Fermi level, which cause a spectral weight redistribution resulting in a much greater DOS at the Fermi level, and a shift upwards of the nodal crossing.  \medskip

\begin{table}[!]
\centering
\caption{Topological characterization of tin pnictides calculated from elementary band representation analysis using \textsc{symtopo} both with and without spin-orbit coupling. }
\label{tin_pnictide_top}
\begin{tabular}{@{}lccc@{}}
\toprule
& \multicolumn{2}{c}{Without SOC} & \multicolumn{1}{c}{With SOC} \\
Material        & Classification & Position & Classification  \\ \midrule
\ce{MgSn2As2} & HSLSM & F-S-$\Gamma$ & TI  \\ 
\ce{CaSn2As2} & HSLSM & F-S-$\Gamma$ & TI \\ 
\ce{SrSn2As2} & HSLSM & F-S-$\Gamma$ & TI  \\
\ce{BaSn2As2} & HSLSM & F-S-$\Gamma$ & TI  \\
\ce{SrSn2P2} & Trivial  & -- & Trivial \\ \bottomrule
\end{tabular}
\end{table}

\noindent
\textbf{Topological characterization}: As the tin pnictides studied here are charge-balanced, the Dirac point is symmetry allowed but dependent on the energy levels and band dispersions.\cite{Gibson2015} Hence, the critical point varies with the constituent elements and the calculation parameters. With the PBE+SOC level of theory, the band inversion resulting in indirect band gaps of \ce{SrSn2As2} and \ce{BaSn2As2} suggest they are topologically nontrivial, whereas the direct band gap without band inversion of \ce{SrSn2P2} suggests it to be a trivial insulator. We screen the topological properties of the five candidate pnictides using the \textsc{symtopo} package which calculates the compatibility conditions of the band representations along high symmetry lines in the Brillouin zone (BZ).\cite{Symtopo} Violations of these conditions indicates a symmetry-protected crossing which is then labelled with the crossing's position in the BZ. Finally, for gapped cases, symmetry-based indicators are used to distinguish topological insulators and topological crystalline insulators. We summarize our results of topological classification in Table \ref{tin_pnictide_top} both with and without SOC. We find all of the Sn-As compounds to be high-symmetry-line semimetals (HSLSMs) without SOC, with the crossing occurring along the F-S-$\Gamma$ high-symmetry line. Including SOC causes the HSLSM to gap out and result in a topological insulator (TI). However, for \ce{SrSn2P2}, both cases with and without SOC result in a trivial phase.

We further investigate the topological properties of a representative Sn-As compound, \ce{SrSn2As2}, by calculating the topological invariants and surface states with WannierTools. We confirm the nontrivial topology by calculating the topological invariant Z\textsubscript{2}. For \ce{SrSn2As2} with PBE+SOC, (\textit{v}\textsubscript{0};\textit{v}\textsubscript{1}\textit{v}\textsubscript{2}\textit{v}\textsubscript{3}) is (1;000) indicating a strong topological insulator. Furthermore, the calculated surface band structure shown in Figure \ref{surface-states} reveals a surface Dirac cone in the bulk band gap at $\Gamma$. We also observe some bending of the surface bands, which is unsurprising due to the small band gap. 

\begin{figure}[!tb]
 \centering
 \includegraphics[width=\columnwidth]{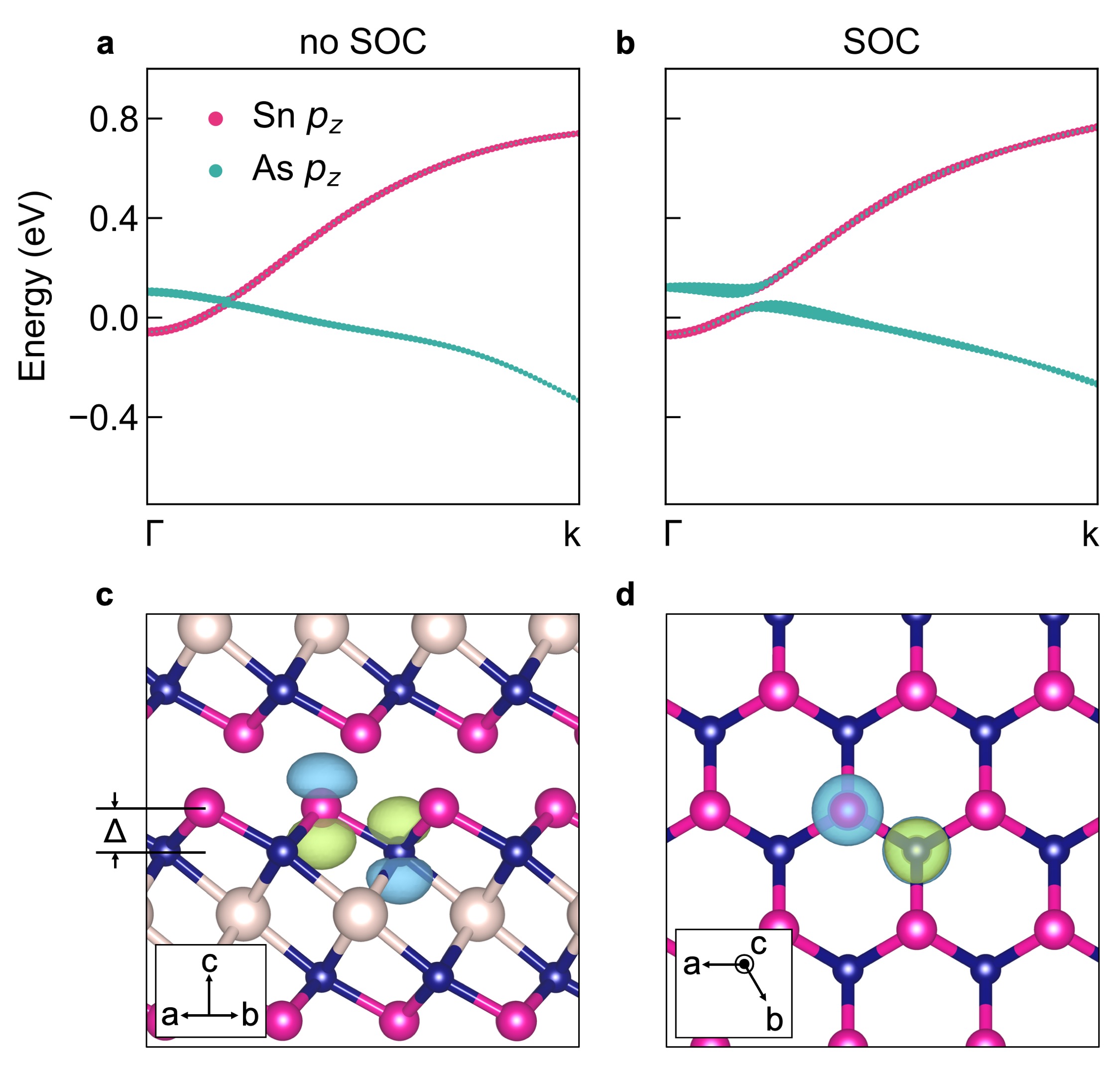}
 \caption{\textbf{a-b:} Zoomed-in section of the \ce{SrSn2As2} band structure where k is a point in the $\Gamma$ to $S_{0}$ direction. The lines are weighted to the orbital character of Sn and As $p_{z}$ states and the Fermi level is set to 0 eV. \textbf{a} Without spin-orbit coupling included there is a band crossing, and \textbf{b} with spin-orbit coupling included there is band inversion and a band gap opening. \textbf{c-d:} Structural details of the Sn--As bilayer  in \ce{SrSn2As2}, with projected Wannier functions of the Sn (dark pink atoms) $p_{z}$ and As (dark blue atoms) $p_{z}$ orbitals. \textbf{c} Side-on view with bilayer height $\Delta$ and \textbf{d} top-down view showing the honeycomb structure.}
 \label{wfs}
\end{figure}

We further analyze a detail of the \ce{SrSn2As2} band structure at a SOC-mediated gap opening close to $E_{F}$ and near $\Gamma$ in Figure \ref{wfs} to elucidate the origins of the critical point. Orbital decomposition of the bands reveals that one is composed of mostly Sn $p_{z}$ states and the other of As $p_{z}$ states. Without SOC (Figure \ref{wfs}a), the bands cross resulting in a Dirac point, whereas when SOC is included (Figure \ref{wfs}b), the bands hybridize, exchange character and a gap opens. Wannier functions of the corresponding Sn $p_{z}$ and As $p_{z}$ orbitals, shown in Figure \ref{wfs}c, reveal that the orbitals align perpendicular to the layers of the structure. A top down view reveals a honeycomb structure within the Sn--As bilayer, Figure \ref{wfs}d, which provides the symmetry protection for the Dirac crossing, as in the Kane—Mele model.\cite{Kane2005} Buckling of the bilayer allows overlap between the $p_{z}$ bonds despite the long Sn--As bond length compared to graphene. As the cation size decreases from Ba to Mg, the buckling of the Sn--As bilayer increases, with a corresponding increase in bilayer height ($\Delta$ indicated in Figure \ref{wfs}c). We find that the bilayer height has a linear relationship with the group velocity at the band crossing shown in Figure \ref{wfs}a, ranging from 3.7~eV{\AA} in \ce{BaSn2As2} to 5.1~eV{\AA} in \ce{MgSn2As2} (SI Table 4).

All three compounds \ce{SrSn2As2}, \ce{BaSn2As2} and \ce{SrSn2P2} are trivial insulators when calculated by HSE06+SOC, as the hybrid functional unwinds the bands past the critical point and the band gap opens up. The dependency of topological order on choice of functional indicates that these compounds lie naturally close to the critical point. This suggests the possibility of tuning the topological order by other degrees of freedom such as strain. For example, as we vary the composition Ca$\rightarrow$Sr$\rightarrow$Ba the HSE06 band gap increases and this coincides with increasing Sn--As bond length. Therefore, the band structure, and hence band gap, can be manipulated towards the critical point by reducing the interlayer distance with a compressive strain.

\subsubsection{\ce{Li6Bi2O7}}
\noindent
\textbf{Structural details}: The structure of \ce{Li6Bi2O7} in the Materials Project database\cite{osti_1298595} is in the space group P2\textsubscript{1}/c (number 14) and consists of \ce{BiO6} octahedra that are both corner- and edge-sharing (shown in Figure \ref{structures}c). Li resides in channels between these octahedra. However, this compound has not been reported by experiment. Upon closer inspection of the composition, we note that the stoichiometry corresponds to Bi being in the +4 oxidation state, although Bi\textsuperscript{+4} is unstable against disproportionation into Bi\textsuperscript{+3} and Bi\textsuperscript{+5} ions.\cite{Mazin1995} Accordingly, a charge-ordered Li\textsubscript{6}Bi\textsuperscript{+3}Bi\textsuperscript{+5}O\textsubscript{7} structure is expected, which would manifest in a difference in bonding on two Bi sites.\cite{Cox1976} The P2\textsubscript{1}/c structure does not capture this, having equivalent bond lengths on every Bi site, with average Bi--O bond length \SI{2.25}{\angstrom}, and equal polyhedral volumes of 14.90 \AA\textsuperscript{3}. We therefore predict that the compound will not exist in the structure given in the database. However, we here examine the given structure and will address the possibility of a lower symmetry, charge-ordered structure in a following work. \medskip

\begin{figure}[!tb]
 \centering
 \includegraphics[width=\columnwidth]{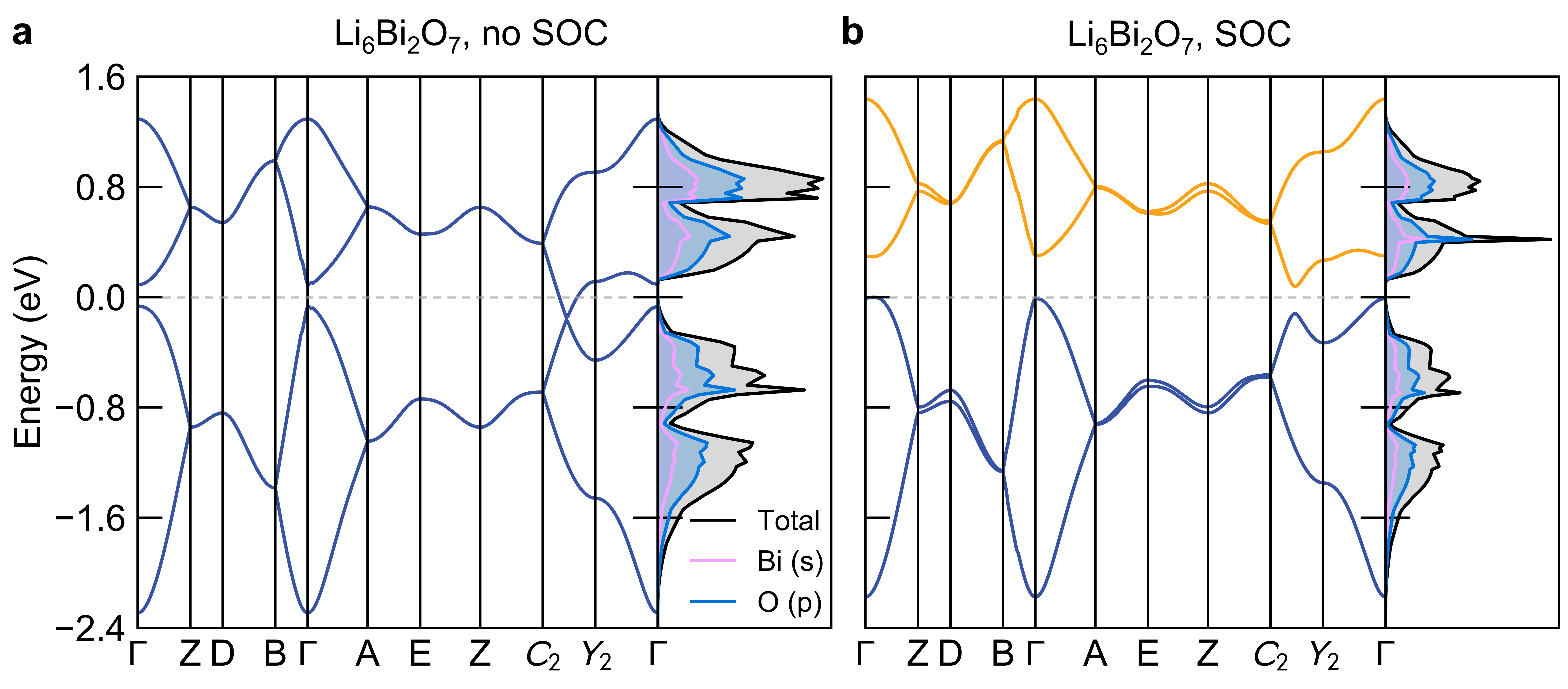}
 \caption{Band structures and orbital-resolved density of states of \ce{Li6Bi2O7} calculated with HSE06: \textbf{a} without including spin-orbit coupling, and \textbf{b} including spin-orbit coupling interactions.}
 \label{li_hse_bs}
\end{figure}

\noindent
\textbf{Electronic structure}: The data set from which the materials were selected predicted that a band gap opened when spin-orbit coupling was included with the PBE functional. However, our PBE calculations which were performed with the converged set of parameters given in SI Table 1 predict the material to be metallic both with and without SOC. When the hybrid functional HSE06 is used, however, a spin-orbit gap does open up. Figure \ref{li_hse_bs}a shows the electronic band structure and orbital-resolved density of states calculated with HSE06 when SOC is not included, showing band crossing just below the Fermi level. Including SOC, Figure \ref{li_hse_bs}b shows a degree of spin splitting and separation of the valence and conduction bands resulting in an indirect band gap of 78 meV. The density of states has a majority O 2\textit{p} and Bi 6\textit{s} character in the region around the Fermi level, indicating that the electronic properties are determined by the hybridization between these two orbitals with the SOC-driven gap resulting from the heavy Bi ion. We conclude that this hypothetical structure indeed fulfils the criteria of a spin-orbit gapped semiconductor, but will not be stable in nature.

\subsection{Suitability as Dark Matter Targets}

The indium chalcogenide compounds \ce{CuInTe2}, \ce{CuInSe2} and \ce{CuInS2} have SOC-induced small band gaps under the generalized gradient approximation. However, the larger band gap predicted by hybrid functionals is in better agreement with the experimentally measured values, as expected for regular semiconductors.\cite{Frick2018c,Neumann1986,Yoshino2001}  Including SOC worsens the prediction of E\textsubscript{g} in comparison to experiment (\ce{CuInTe2} has an optical band gap of 0.9 eV; HSE06 predicts 0.74 eV without SOC and 0.54 eV with SOC), but the effects are needed to reveal features of the experimentally measured band structure such as splitting of the valence band.\cite{Frick2018c,Neumann1986} Regardless, the band gaps of the compounds have been comprehensively determined by experiment, and, being of the order of an eV, are too large for our desired absorption of DM with meV masses or scattering of DM with keV masses.

The tin pnictides have an electronic structure that is strongly influenced by SOC due to the topological nature of the compounds. At the PBE level of theory, the structures exhibited band inversion such that even when the hybrid functional was used, the band gaps opened up as expected but remained at the meV scale. With this change in functional, \ce{SrSn2As2} and \ce{BaSn2As2} pass through the topological critical point, resulting in direct band gaps of 70 and 112 meV respectively. \ce{SrSn2P2} is a trivial insulator in both cases, and the change in functional causes a larger gap opening to 198 meV. These three materials can offer improved sensitivity to light dark matter interactions over traditional semiconducting compounds with eV scale band gaps. The 70-200 meV range predicted by HSE06 provides multiple options for precisely targeted and tunable light DM masses. Additionally, as the composition is varied by changing the cation, the Sn--As bilayer height is modulated which has the effect of also tuning the group velocity. For \ce{CaSn2As2}, \ce{SrSn2As2} and \ce{BaSn2As2}, this band crossing is close to the Fermi level, such that the Fermi velocity ($v_F$) is varied from 4.3 to 3.7 eV\AA. To maximize the DM-scattering rate, $v_F$ should be the same as the velocity of DM, which is $\sim10^{-3}c$.\cite{Hochberg2018} The tin pnictides are close to an ideal match: \ce{CaSn2As2} $v_F$ = $3.5\times10^{-4}c$, \ce{SrSn2As2} $v_F$ = $3.1\times10^{-4}c$, \ce{BaSn2As2} $v_F$ = $3.0\times10^{-4}c$. These values are comparable to \ce{ZrTe5} ($v_{F,x}$ = $2.9\times10^{-3}c$, $v_{F,y}$ = $5.0\times10^{-4}c$, $v_{F,z}$ = $2.1\times10^{-3}c$) which provides an excellent DM reach.\cite{Hochberg2018} Like \ce{ZrTe5}, the tin pnictides also have anisotropic velocities, which will enable directional detection for capturing daily or annual modulation of a DM signal, distinguishing it uniquely from background signals. For example, the group velocities vary by two orders of magnitude: $v_{g,x,y}$ = $3.1\times10^{-4}c$ and $v_{g,z}$ = $2.0\times10^{-6}c$ for \ce{SrSn2As2}.

The band gap and Fermi velocity of the target material determine the lower bound of the mass sensitivity and its cross section with DM respectively. Anisotropic Fermi velocities can provide directional targets whereby the incoming DM wind gives a directional dependence -- the `smoking gun' of DM detection. Therefore, being able to tune these two critical parameters within one family of materials offers substantial benefits for the design of detection experiments. Furthermore, there is the possibility to expand this range further than considered here by varying the composition through different combinations of cation and pnictogen. Crucially,  solid solutions have already been experimentally realized, and mixtures on either cation or anion sites could provide fine tuning of the band gap.\cite{Asbrand1995} In fact, for certain DM interactions a direct band gap is preferred over an indirect band gap, as the higher probability of a direct excitation improves the target sensitivity.

The \ce{Li6Bi2O7} compound also has an meV scale band gap predicted by HSE06+SOC, however the structure examined here is likely to be unstable and any symmetry lowering is likely to affect the band gap. More specifically, charge disproportionation is likely to lead to a lowering of the symmetry of the structure, leading to a reduction in orbital overlap and increase of the band gap. This compound has not been reported previously, and there are no experimental measurements available.

\subsection{Theoretical Predictions of Spin-Orbit Gapped Materials}

The analysis presented here highlights some of the drawbacks of a DFT-based search for low band gap materials. Of the three materials families which had a spin-orbit gap predicted by PBE, two did not maintain meV-scale band gaps under closer investigation. Some of these shortcomings have been discussed previously in relation to predicted topological materials,\cite{Vidal2011,Vergniory2013,Zunger2019,Malyi2020} however there is no one method that will give reliable predictions in all cases.

The most important consideration for detector applications is the structural stability, and hence experimental realization of the predicted target. If a compound has not previously been synthesized, then its thermodynamic stability can be estimated by phonon analysis or energy above the convex hull, if necessary identifying competing phases using structure prediction tools. In the case of \ce{Li6Bi2O7}, even though a stable energy above hull was predicted\cite{osti_1298595} (within calculation error), the insufficient treatment of charge localization by PBE failed to capture a structural distortion stemming from charge disproportionation. While the prediction of experimentally unfeasible structures is not unique to the field of topological materials, the symmetry requirements of non-trivial topology mean that any symmetry-breaking structural changes can invalidate a topological analysis. More generally, the close structure-property relationship of spin-orbit band gaps implies that even small changes in geometry will have a large effect.

A second drawback of DFT-led searches for low band gap materials is the underestimation of trivial band gaps by standard DFT approximations; this has the opposite effect of overestimation of gaps with band inversion, which can lead to false positive topological materials and incorrect band dispersions.\cite{An2014,Forster2015,Forster2016} This has remained a caveat for high-throughput computational searches of topological insulators.\cite{Yang2012a, Olsen2019} Going beyond DFT, the GW approximation is the most accurate method for predicting electronic properties without parameterization, improving upon both the bulk and surface electronic structure of topological insulators and giving results consistent with experimental photoemission, optical and EELS spectra.\cite{Aguilera2015, Aguilera2019} However, quasiparticle self-consistent GW (QSGW) is known to systematically overestimate band gaps,\cite{Svane2011} which can lead to false negative topological classifications of small band gap materials. Furthermore, given the many variations of GW available, electronic properties are sensitive to computational choices such as the number of self-consistent steps, whether SOC is included directly or as a perturbation, and the ad-hoc correction of the hybrid QSGW scheme. These can cause gap variations greater than 1 eV.\cite{Garza2016} For materials close to the topological critical point, these choices can result in qualitatively different topological classifications.\cite{Svane2011,Aguilera2013,Aguilera2013a}

Importantly, depending on the system size and properties of interest, calculations with GW approximations can be prohibitively expensive. The hybrid density functionals, which eliminate much of the self-interaction error of DFT by including a fraction of exact Hartree-Fock exchange, have been extremely successful as a mid-cost level of theory that can give GW-quality results for topological materials.\cite{Vidal2011,Crowley2015,Malyi2020} However, the screened hybrid functionals, such as the HSE06 functional used here, rely on fixed parameters for screening length and percentage of exact exchange, and predicted band gaps are dependent on these parameters. A single hybrid functional with a fixed amount of exact exchange cannot accurately describe small and large band gap materials simultaneously, and the settings that have been benchmarked for general use will tend to overestimate the band gaps of narrow-gapped materials.\cite{Garza2016} In topological insulators, this could lead to underestimations of inverted band gaps and false negative topological classifications. Therefore, if parameterization via experimental results is not possible, hybrid functionals are not necessarily more accurate than PBE.\cite{Li2014a} In the case of the topological semimetal GaGeTe, HSE overestimated the band gap and predicted a trivial gap, whilst PBE gave a closer match to the measured band gap and supported the topological classification from experiment, although the nature of the indirect gap was not captured by either HSE or PBE.\cite{Haubold2019} The family of tin pnictides shown here is one example of a topological classification that is dependent on the choice of PBE or hybrid exchange correlation functional, and there are other examples in literature.\cite{Vidal2011,Sun2011b} Aside from the treatment of charge localization, even small differences in geometry from different functionals can affect the topological classification.\cite{Reid2020} In these edge cases, careful consideration of the electronic structure can be taken on a case-by-case basis, however, experimental verification is always necessary.

\section*{Conclusions}

Three materials, \ce{CuInTe2}, \ce{SrSn2As2} and \ce{Li6Bi2O7}, predicted by DFT to have band gaps induced by spin-orbit coupling interactions, were investigated for their electronic and topological properties. These materials and a range of isostructural compounds were evaluated for their suitability as low-mass dark matter detection targets.

The band gaps of \ce{CuInTe2}, \ce{CuInSe2} and \ce{CuInS2} predicted by HSE06+SOC were found to be in good agreement with experiment, but are too large to be sensitive to light dark matter. Likewise with \ce{Li6Bi2O7}, using HSE06+SOC led to an increased prediction of the band gap compared to PBE, but structural distortions associated with charge disproportionation must be further investigated, and synthesis routes explored.

The family of tin pnictides, however, has several properties making them promising as targets for light DM detection. Firstly, direct band gaps ranging from 70-200 meV are predicted across the three compounds \ce{SrSn2As2}, \ce{BaSn2As2} and \ce{SrSn2P2} and can be tuned by alloying, making them sensitive to sub-GeV DM candidates. Secondly, the tunable Fermi velocity suggests that these compounds can be kinematically matched with DM to optimize the cross section between DM and electrons in the target, additionally providing a route for directional direction. Finally, some of the family have already been synthesized in crystal form and found to lie close to a topological critical point. However, further experimental studies are needed to fully characterize the structure-property phase space in addition to further variations on composition for tuning of the band gap both within and beyond the range presented here. 

\section*{Acknowledgements}
We thank Anubhav Jain for discussions and help generating the spin-orbit coupling data set used by this work. We also thank Junsoo Park for helpful discussions regarding Wannier functions. S.G. and K.I. were supported by the Laboratory Directed Research and Development Program of LBNL under the U.S. Department of Energy (DoE) Contract No. DE-AC02-05CH11231. Computational resources were provided by the National Energy Research Scientific Computing Center and the Molecular Foundry, DoE Office of Science User Facilities supported by the Office of Science of the U.S. Department of Energy under Contract No. DE-AC02-05CH11231. The work performed at the Molecular Foundry was supported by the Office of Science, Office of Basic Energy Sciences, of the U.S. Department of Energy under the same contract. A.F. was funded by the DOE Basic Energy Sciences program—the Materials Project—under Grant No. KC23MP.

\bibliography{refs}
\end{document}


\title{Supporting Information: Prediction of Tunable Spin-Orbit Gapped Materials for Dark Matter Detection}

\author{Katherine Inzani}
\email{kinzani@lbl.gov}
\affiliation{Materials Science Division, Lawrence Berkeley National Laboratory, Berkeley, CA 94720, USA}
\affiliation{Molecular Foundry, Lawrence Berkeley National Laboratory, Berkeley, CA 94720, USA}
\author{Alireza Faghaninia}
\affiliation{Energy Technologies Area, Lawrence Berkeley National Laboratory, Berkeley, CA 94720, USA}

\author{Sin\'{e}ad M. Griffin}
\affiliation{Materials Science Division, Lawrence Berkeley National Laboratory, Berkeley, CA 94720, USA}
\affiliation{Molecular Foundry, Lawrence Berkeley National Laboratory, Berkeley, CA 94720, USA}

\date{\today}
{\let\clearpage\relax\maketitle}

\begin{table}[h!]
\centering
\begin{threeparttable}
\captionsetup{justification=raggedright}
\caption{Energy cutoffs and k-point grids used for calculations on the primitive cell of each of the reported materials.}
\label{my-label}
\begin{tabular}{@{}lccc@{}}
\toprule
Material       & xc-functional & Energy cut-off (eV) & k-point grid \\ \midrule
\ce{InCuTe2}   & PBE       & 600  & $4\times4\times4$    \\
               & HSE06     & 600  & $4\times4\times4$    \\
\ce{InCuSe2}   & PBE       & 600  & $4\times4\times4$    \\
               & HSE06     & 600  & $4\times4\times4$    \\
\ce{InCuS2}    & PBE       & 600  & $4\times4\times4$    \\
               & HSE06     & 600  & $4\times4\times4$    \\
\ce{SrSn2As2}  & PBE       & 400  & $12\times12\times12$ \\
               & HSE06     & 400  & $6\times6\times6$    \\
\ce{MgSn2As2}  & PBE       & 400  & $12\times12\times12$ \\
\ce{CaSn2As2}  & PBE       & 400  & $12\times12\times12$ \\
               & HSE06     & 400  & $5\times5\times5$    \\
\ce{BaSn2As2}  & PBE       & 400  & $12\times12\times12$ \\
               & HSE06     & 400  & $9\times9\times9$    \\
\ce{SrSn2P2}   & PBE       & 400  & $12\times12\times12$ \\
               & HSE06     & 400  & $4\times4\times4$    \\
\ce{Li6Bi2O7}  & PBE       & 650  & $2\times2\times2$    \\ 
               & HSE06     & 650  & $2\times2\times2$    \\ \bottomrule
\end{tabular}
\end{threeparttable}
\end{table}

\begin{table}[h!]
\centering
\begin{threeparttable}
\captionsetup{justification=raggedright}
\caption{Optimized lattice parameters of copper indium chalcogenides and deviation from the reported experimental values.}
\label{my-label1}
\begin{tabular}{@{}lccccccc@{}}
\toprule
& \multicolumn{2}{c}{Calculated (\AA)} & \multicolumn{2}{c}{Experimental (\AA)} & \multicolumn{2}{c}{Difference (\%)} & \\
Material        & a & c & a & c & $\Delta$a & $\Delta$c & Reference \\ \midrule
\ce{CuInTe2} & 6.291 & 12.647 & 6.204 & 12.404 & -1.41 & -1.96 & \citenum{Frick2018h} \\ 
\ce{CuInSe2} & 5.869 & 11.801 & 5.873 & 11.583 & 0.07 & -1.88 & \citenum{Rincon1992} \\ 
\ce{CuInS2} & 5.573 & 11.226 & 5.517 & 11.122 & -1.01 & -0.94 & \citenum{Hwang1978} \\ \bottomrule
\end{tabular}
\end{threeparttable}
\end{table}

\begin{figure}[h!]
 \centering
 \captionsetup{width=16cm,justification=raggedright}
 \includegraphics[width=16cm]{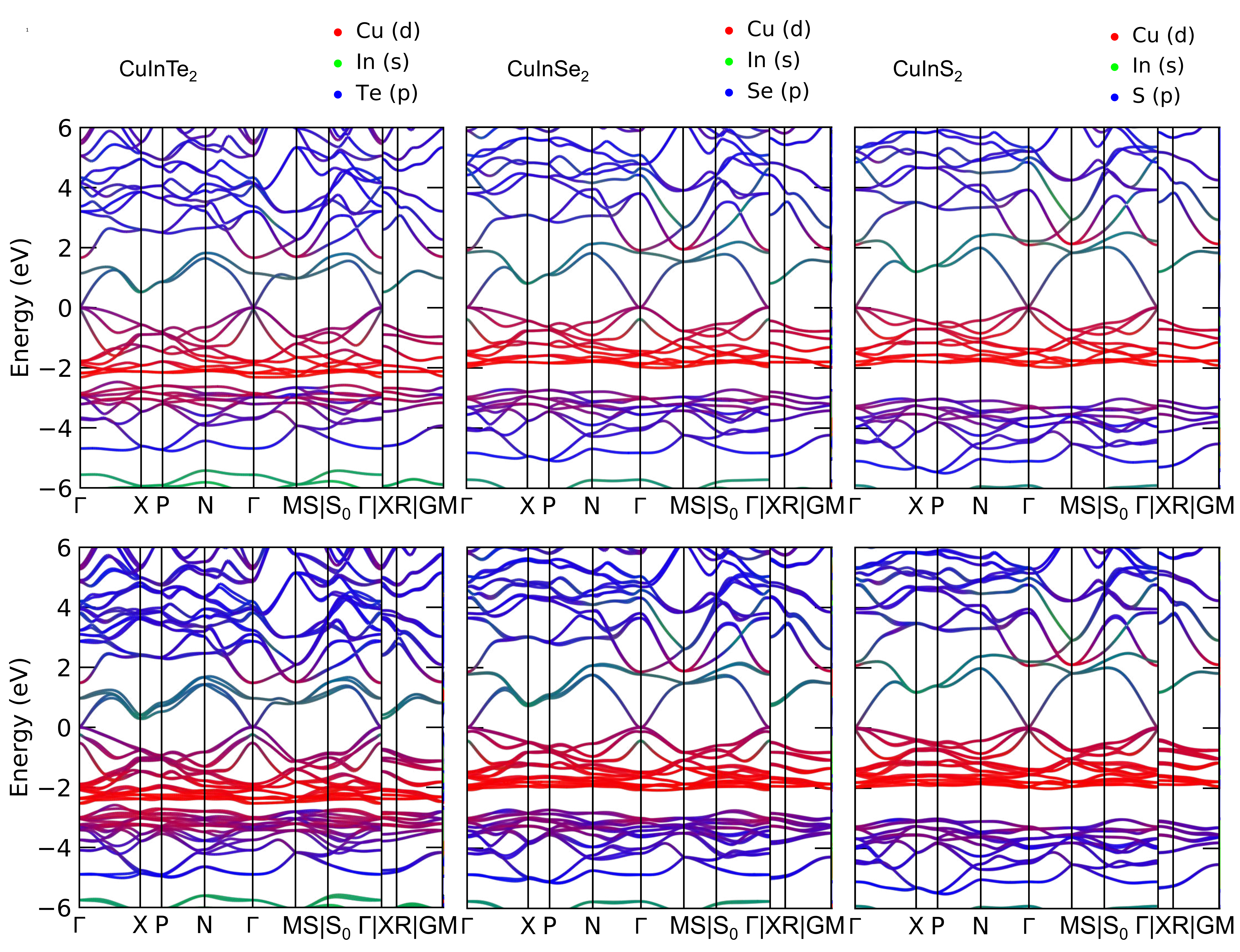}
 \caption{Orbital projected band structures of \ce{CuInTe2}, \ce{CuInSe2} and \ce{CuInS2} calculated with PBE. The top row is without spin-orbit coupling included and the bottom row is with spin-orbit coupling included.}
 \label{cuin_pbe_bs}
\end{figure}

\begin{figure}[h!]
 \centering
 \includegraphics[width=16cm]{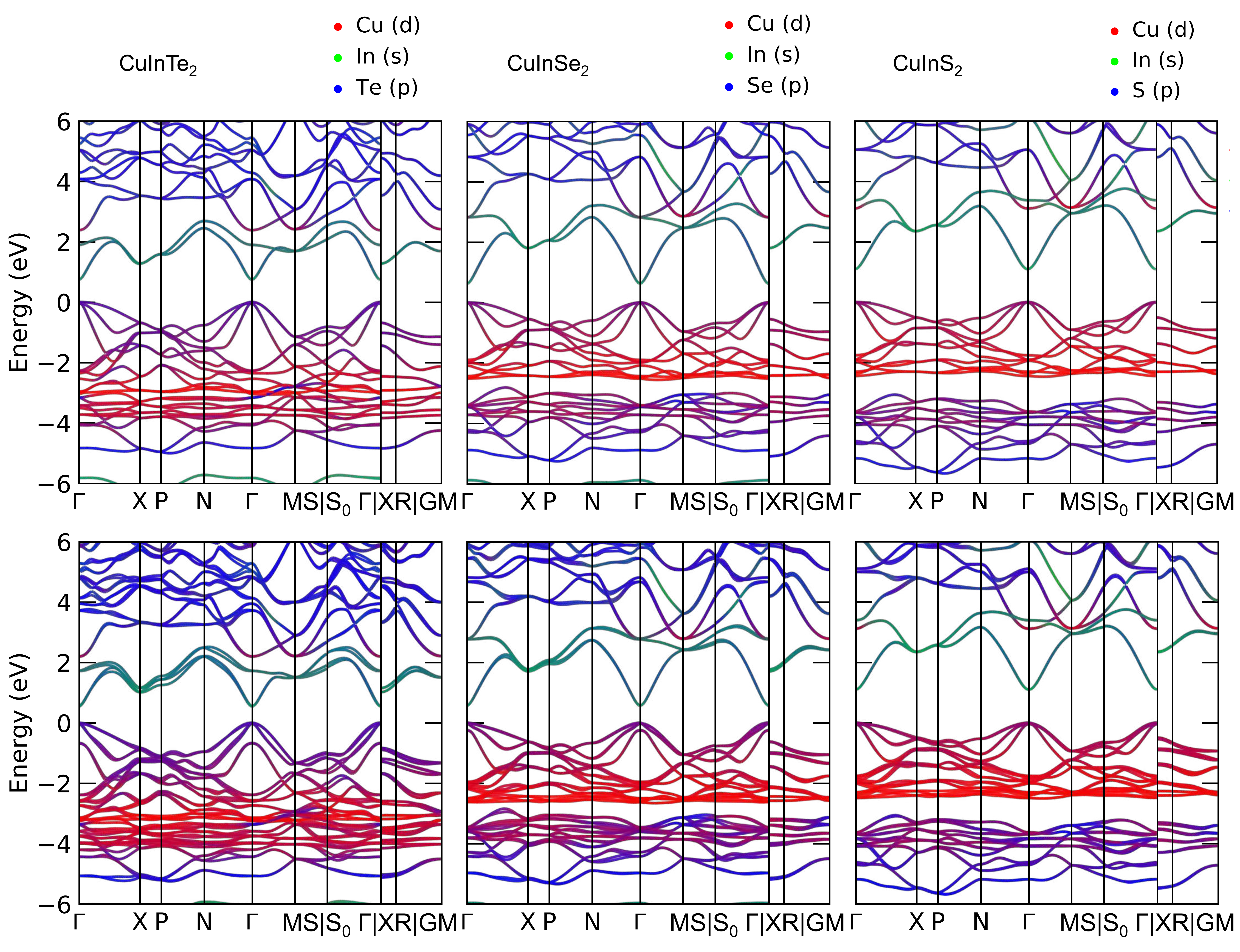}
 \captionsetup{width=16cm,justification=raggedright}
 \caption{Orbital projected band structures of \ce{CuInTe2}, \ce{CuInSe2} and \ce{CuInS2} calculated with HSE06. The top row is without spin-orbit coupling included and the bottom row is with spin-orbit coupling included.}
 \label{cuin_hse_bs}
\end{figure}

\begin{table}[h!]
\centering
\begin{threeparttable}
\captionsetup{justification=raggedright}
\caption{Optimized lattice parameters of tin pnictides and deviation from the available reported experimental values in Reference \citenum{Asbrand1995}.}
\label{my-label2}
\begin{tabular}{@{}lcccccc@{}}
\toprule
& \multicolumn{2}{c}{Calculated (\AA)} & \multicolumn{2}{c}{Experimental (\AA)} & \multicolumn{2}{c}{Difference (\%)} \\
Material        & a & c & a & c & $\Delta$a & $\Delta$c \\ \midrule
\ce{MgSn2As2} & 4.092 & 25.657 & N/A & N/A & -- & -- \\ 
\ce{CaSn2As2} & 4.205 & 26.583 & 4.142\tnote{*} & 26.017\tnote{*} & -1.54 & -2.18 \\ 
\ce{SrSn2As2} & 4.276 & 27.248 & 4.204 & 26.726 & -1.71 & -1.95 \\
\ce{BaSn2As2} & 4.350 & 28.021 & 4.221\tnote{**} & 28.464\tnote{**} & -3.04 & 1.55 \\
\ce{SrSn2P2} & 4.147 & 26.940 & N/A & N/A & -- & -- \\ \bottomrule
\end{tabular}
\begin{tablenotes}\footnotesize
\item[*]Value extrapolated from \ce{NaSn2As2} and Na\textsubscript{0.3}Ca\textsubscript{0.7}Sn\textsubscript{2}As\textsubscript{2}.
\item[**]Value extrapolated from \ce{NaSn2As2} and Na\textsubscript{0.6}Ba\textsubscript{0.4}Sn\textsubscript{2}As\textsubscript{2}.
\end{tablenotes}
\end{threeparttable}
\end{table}

\begin{figure}[h!]
 \centering
 \includegraphics[width=16cm]{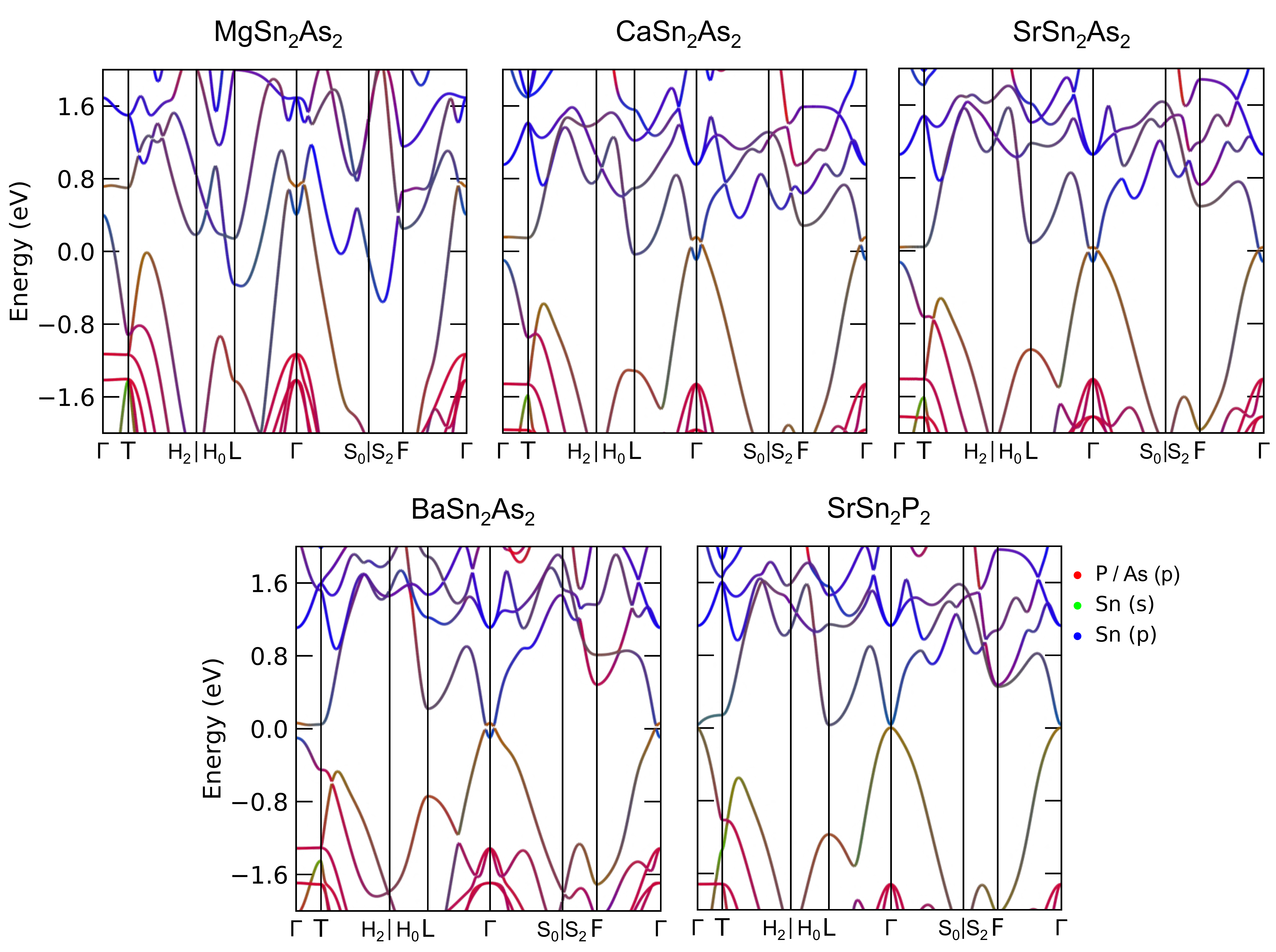}
 \captionsetup{width=16cm,justification=raggedright}
 \caption{Orbital projected band structures of \ce{MgSn2As2}, \ce{CaSn2As2}, \ce{SrSn2As2}, \ce{BaSn2As2} and \ce{SrSn2P2} calculated by PBE without spin-orbit coupling included.}
 \label{tin_gga_cl_bs}
\end{figure}

\begin{figure}[!tb]
 \centering
 \includegraphics[width=14.2cm]{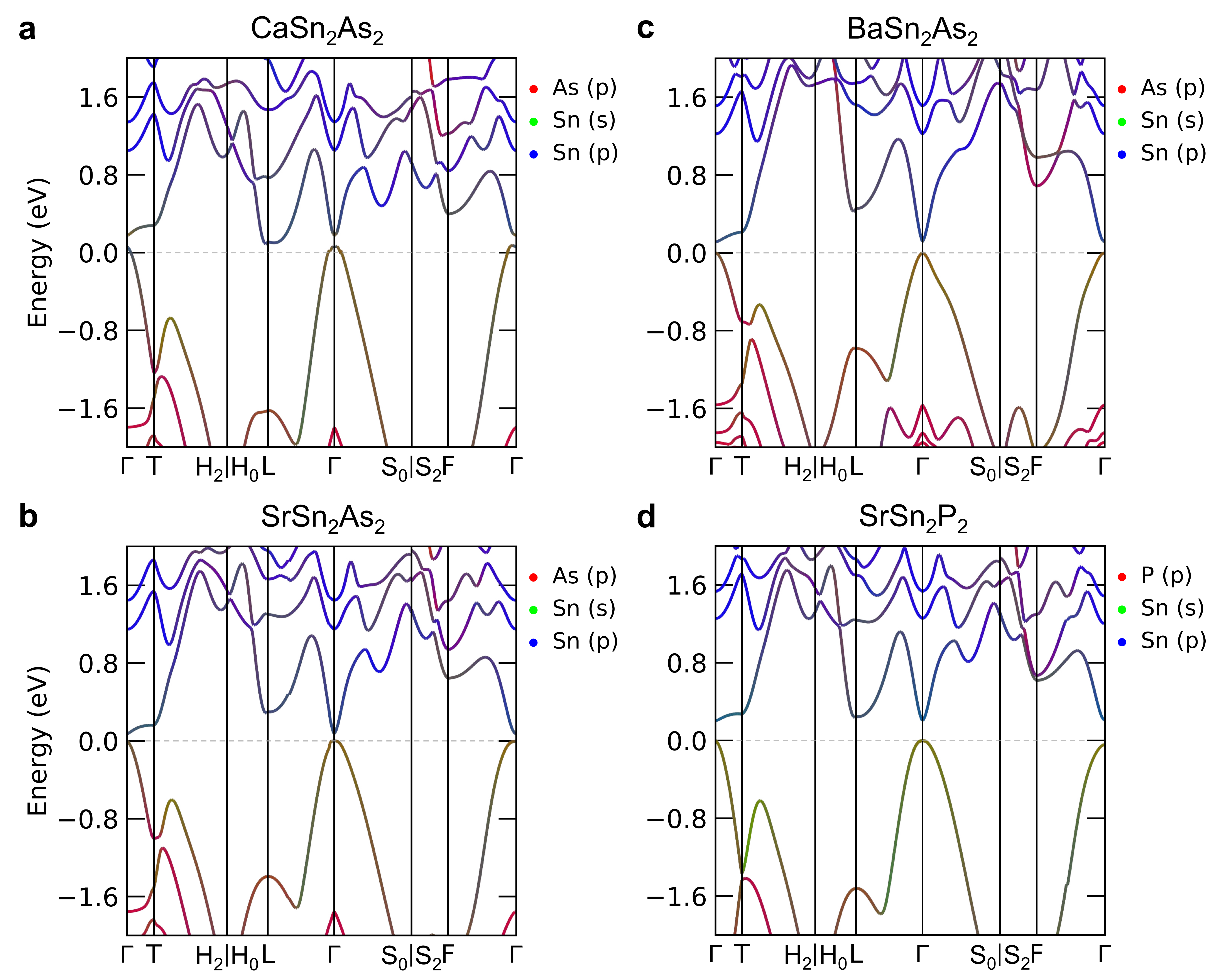}
 \captionsetup{width=14.2cm,justification=raggedright}
 \caption{Orbital projected band structures of tin pnictides calculated by HSE06 with spin-orbit coupling included: \textbf{a} \ce{CaSn2As2}, \textbf{b} \ce{SrSn2As2}, \textbf{c} \ce{BaSn2As2} and \textbf{d} \ce{SrSn2P2}.}
 \label{tin_hse_ncl_bs}
\end{figure}

\begin{figure}[!tb]
 \centering
 \includegraphics[width=14cm]{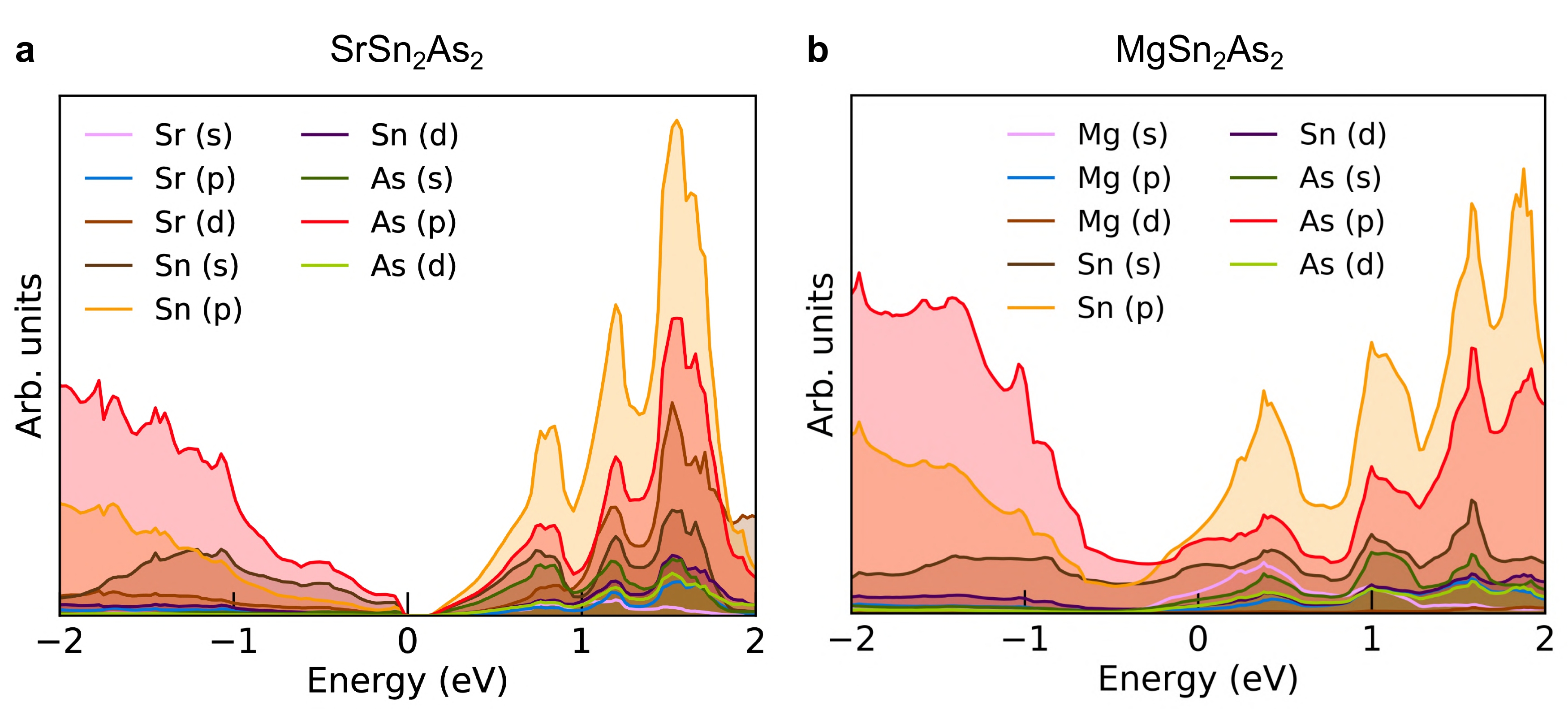}
 \captionsetup{width=14cm,justification=raggedright}
 \caption{Orbital-resolved density of states of \textbf{a} \ce{SrSnAs2} and \textbf{b} \ce{MgSn2As2}.}
 \label{srsn2as2_dos}
\end{figure}

\begin{table}[h!]
\centering
\begin{threeparttable}
\captionsetup{justification=raggedright}
\caption{Trend in Fermi velocity ($v_{F}$) with geometry in the tin pnictides. The slope of the band composed of Sn \textit{p} states is taken at the band crossing along the $\Gamma$ to $S_{0}$ direction from the band structures calculated by PBE with no spin-orbit coupling.}
\label{bandslopes}
\begin{tabular}{@{}lcccccc@{}}
\toprule
Material       & \thead{Sn--As\\bond length (\AA)}   &   \thead{Sn--As\\bilayer height (\AA)}   &   \thead{$v_{F}$ (eV\AA)} \\ \midrule
\ce{MgSn2As2} &  2.796   &   1.496    &    5.1  \\
\ce{CaSn2As2} &  2.809   &   1.412    &    4.3  \\
\ce{SrSn2As2} &  2.826   &   1.376    &    3.9  \\
\ce{BaSn2As2} &  2.847   &   1.341    &    3.7  \\ \bottomrule
\end{tabular}
\end{threeparttable}
\end{table}

\FloatBarrier
\bibliography{SI_refs}